\documentclass[pra,lengthcheck,twocolumn,showpacs,amssymb]{revtex4}
\usepackage{amsmath,bbm,epsfig}
\usepackage{amsfonts}
\usepackage{mathrsfs,psfrag}
\usepackage{graphicx}
\newcommand{\epl}{Europhys.\ Lett.\ }

\newcommand{\jpa}{J.\ Phys.\ A\ }

\newcommand{\jpf}{J.\ Phys.\ France\ }

\newcommand{\ii}{\mbox{i}}
\newcommand{\ket}[1]{\ensuremath{| #1 \rangle}}
\newcommand{\bra}[1]{\ensuremath{\langle #1 |}}

\begin{document} 

\title{Engineering many-body quantum dynamics by disorder}

\author{Pierfrancesco Buonsante$^1$ and Sandro Wimberger$^{1,2}$}  
\affiliation{
$^1$CNISM -- Dipartimento di Fisica, Politecnico di Torino, Corso Duca 
degli Abruzzi 24, 10129 Torino, Italy \\
$^2$Institut f\"ur Theoretische Physik, Universit\"at Heidelberg,
Philosophenweg 19, 69120 Heidelberg, Germany
}

\date{\today}

\begin{abstract}
Going beyond the currently investigated regimes in experiments on quantum transport of 
ultracold atoms in disordered potentials, we predict a crossover between regular and 
quantum-chaotic dynamics when varying the strength of disorder. Our spectral approach is based
on the Bose-Hubbard model describing interacting atoms in deep random potentials.
The predicted crossover from localized to diffusive dynamics depends on the {\it simultaneous} 
presence of interactions and disorder, and can be verified in the laboratory by monitoring
the evolution of typical experimental initial states.
\end{abstract}


\maketitle

While well-controlled experiments in solid-state systems are lacking,
the recent advances in atom and quantum optics allow the experimentalist to 
study minimal models where single-particle dynamics, many-body interactions, and disorder
can be engineered at will. Ultracold bosons or fermions loaded into optical lattices, 
which realize spatially periodic potentials \cite{Bloch2005}, 
are optimally suited to study, e.g., quantum transport across disorder potentials and possible manifestations 
of Anderson localization in the mean-field regime \cite{Lye2005,Paul2007}. Moreover, modern experiments
reach the regime of strong atom-atom correlations to investigate many-body quantum effects such as 
interaction-driven phase transitions \cite{Bloch2005,BEC_mott} or interaction-induced changes of Landau-Zener 
tunneling rates \cite{Ess2005,TMW2007}.

The Bose-Hubbard model well describes ultracold bosons in periodic lattices at small fillings
(where a mean-field theory is obviously bound to fail) and not too shallow lattice depths 
\cite{Bloch2005,BEC_mott,BK2003,KB2004}. 
A recent study of an open Bose-Hubbard system \cite{TMW2007} furthermore showed that 
many-body interactions lead to similar decay-rate distributions as 
predicted for single-particle transport in disordered potentials. 
More precisely, interactions in a many-body system can substitute for disorder in the 
diffusive regime of quantum transport \cite{TMW2007}. 

Here we show how to engineer the dynamical properties of a many-body Bose-Hubbard system by
varying the strength of static disorder. 
We predict that, for an intermediate regime of disorder strength, the system
shows clear signature of {\it global} quantum chaos. The latter is quantified by spectral measures
of quantum chaos \cite{WPI1997,BK2003,KB2004,TMW2007,MPBS1993} and transport \cite{CGIMZ1994,KM1993,RB2003}. 
Complexity arises in our systems from the {\it simultaneous} presence of atom-atom interactions 
and disorder. 

We consider a disordered Bose-Hubbard system on a 1D lattice \cite{SBZ1991} comprising $L$ sites, defined by the Hamiltonian
\begin{equation}
H \!=\! \sum_{\ell=1}^{L} \! \left[ U\, a_\ell^\dag\,^2 \, a_\ell^2 \! -\! J \left(e^{i \theta} a_\ell^\dag a_{\ell+1}
+ h.c. \right) \!+\!\epsilon_\ell a_\ell^\dag a_\ell \right] \!.
\label{eq:1} 
\end{equation}
The operators $a_\ell^\dag$, $a_\ell$ create and destroy bosons at lattice site $\ell$, respectively. The random on-site potentials, $\{\epsilon_\ell\}$, are chosen from a box distribution in $[-\epsilon/2,\epsilon/2]$. The deterministic Peierls phase $\theta$ in the kinetic term corresponds to a finite (angular) momentum of
the lattice or, equivalently, to the presence of a (effective) magnetic potential \cite{phase}, and mathematically to imposing different boundary conditions, {\it twisted} as opposed to simply periodic, onto the standard model, Eq.~(\ref{eq:1}) with $\theta=0$ \cite{RB2003}.  Phases could be controlled experimentally as described in \cite{AOC2005}, while the periodic boundary conditions assumed for (\ref{eq:1}) in the following could be implemented in optical ring lattices \cite{MG2004,AOC2005}. The effects of $\{\epsilon_\ell\}$ and $\theta$ are in some sense complementary. This can be understood switching to the Fock basis of the quasimomentum (QM) 
operators $b_\kappa = L^{-1/2} \sum_\ell  a_\ell e^{{\rm i}2 \pi \ell \kappa /L}$ \cite{KB2004,BPV2005}, diagonalizing the kinetic term in (\ref{eq:1}).  In this reciprocal basis, the interaction term is block-diagonal, the blocks being labeled by the total QM$=\sum_\kappa \kappa b_\kappa^\dag b_\kappa {\rm \,mod}(L)$ -- where the $\rm mod$ operation guarantees a QM in the unit interval.  Conversely, the local potential term, $\{\epsilon_\ell\}$, couples blocks of different QM, since it corresponds to a sum of nonlocal operators $b_\kappa^\dag b_\eta$ \cite{BPV2005}. Any such operator couples blocks whose total QM differs by $(\kappa -\eta)$ units, and since $b_\kappa^\dag b_\kappa$ is a number operator, the coupling within each block reduces to a trivial constant. 
Hence, $\{\epsilon_\ell\}$ induces interblock couplings, whilst $\theta$ affects the diagonal blocks of fixed QM.

We are interested in the global dynamics generated by \eqref{eq:1}. Our approach to characterize
the quantum transport in the system is twofold: first, we study the spectral properties of \eqref{eq:1}, and
secondly, we present results of the time-evolution of initial states which are not eigenstates of
\eqref{eq:1}. The  evolution  of  typical experimental observables, 
such as the spatial population and the mean momentum of the condensate particles
\cite{Bloch2005,BEC_mott,Lye2005}, allows one to directly probe the here predicted diffusion-localization transition (DLT). 
In contrast to \cite{K2006}, we do 
not just focus on the regime of small disorder $\epsilon \ll J$, and we include $\theta$ to 
generalize the assumed periodic boundary conditions.

For the case without disorder and filling factors of order one, a cross-over between regular 
and quantum chaotic spectra was predicted in \cite{KB2004} when varying the ratio
$U/J$. Quantum chaotic spectra are identified by their statistics, more precisely the distribution
$P(s)$ of the normalized level spacings $s\equiv \Delta E/ \overline{\Delta E}$ follows a Wigner-Dyson
(WD) distribution \cite{KB2004}. 
Also by adding a linear force to the Hamiltonian, a transition between regular and chaotic motion 
can be identified by a cross-over from Poisson to WD statistics \cite{BK2003,TMW2007}, corresponding to
a regime of strong (Stark) localization or of quantum chaos, respectively. 
Based on translational invariance, most previous results \cite{BK2003,KB2004,TMW2007} concern a 
subset of states with constant (conserved) QM. Since in the experiment it is generally hard to 
focus on a subset of states taking part in the dynamical evolution, it is desirable and more 
general to search for {\it global} quantum chaos, which is not just restricted to 
independent subsets of the system's eigenvalue spectrum, e.g. corresponding to constant QM. 
Disorder (which cannot be completely avoided in any real system) 
naturally breaks the translation invariance \cite{K2006}, and we will show that the 
dynamics of a dilute boson system (induced by its spectral properties) can be controlled 
by the combined action of atom-atom interactions and the random potential in \eqref{eq:1}.

\begin{figure}
  \centerline{\epsfig{figure=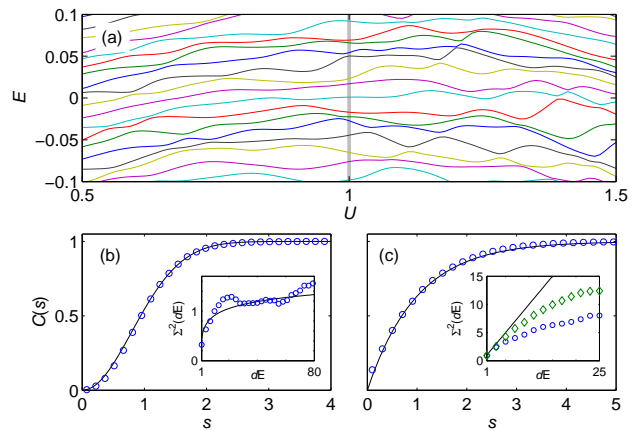,width=\linewidth}} 
  \caption{(color online) (a) 
           Sector of the spectrum in the band center as a function of $U$, for 
           $N=3,L=15,J=1,\theta \simeq 0.119$ for $\epsilon=1$ 
           (a linear function of $U$ was added to $E$ to eliminate an overall trend).
           (b) $C(s)$ (obtained from collecting levels of 25 disorder realizations) 
           and $\Sigma^2$ (for one realization in the inset) at $U=1$, corresponding to the WD
           prediction (solid lines). (c) same as in b) but for a regular case with
           $N=3,L=15$ ($\circ$), $20$ ($\diamond$),
           $U=1=J$ and $\epsilon=10, \theta=0$, together with the Poisson predictions (solid lines).
   }
  \label{fig:1}
\end{figure}

For a complex spreading of the system's eigenfunctions in the eigenbases of the integrable 
cases ($J=0=\epsilon$) and ($U=0=\epsilon$), the energy scales defined by the terms in \eqref{eq:1} 
should be roughly of the same order of magnitude, i.e., $J \sim U \sim \epsilon$. 
For such a situation and small filling, we indeed found clear signatures
of {\it global} quantum chaos in the system. Our results are shown in Fig.~\ref{fig:1}, 
which collects avoided crossing scenarios of the energy levels (a), the
cumulative distribution $C(s) \equiv \int_0^{s}ds'\,P(s')$, and the number variance $\Sigma^2$
of the energy levels (which measures the long-range correlations in the spectrum \cite{MPBS1993}) in (b,c).
Both $C(s)$ and $\Sigma^2$ well agree with the WD predictions for a Gaussian Orthogonal Ensemble of 
random matrices \cite{WPI1997,BK2003,KB2004,MPBS1993}. 
Residual symmetries of \eqref{eq:1} (e.g., a reflection symmetry for special values of the 
number of atoms $N$ and the system size $L$ \cite{KB2004})
can be destroyed by $\theta \neq 0$. 

\begin{figure}
  \centerline{\epsfig{figure=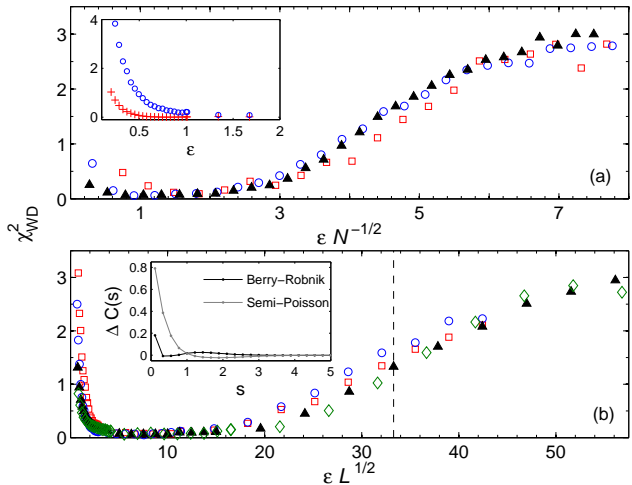,width=\linewidth}} 
  \caption{(color online) (a,b) $\chi^2$ statistical test \cite{nr}
           (with values close to zero for good WD statistics) as a function of the
           {\it scaled} disorder parameter, for (a) $L=15$ and
           $N=2$ ($\Box$), $3$ ($\circ$), $4$ ($\blacktriangle$), and (b) for $N=4$ and $L=8$ ($\Box$), 
           $10$ ($\circ$), $14$ ($\blacktriangle$), $17$ ($\diamond$), at $J=1=U, \theta=0$.
           The inset in (a) shows a zoom of $N=3$ for small $\epsilon$, $\theta =0$ ($\circ$)
           and $\theta \simeq 0.119$ ($+$). Each data point is averaged over 20 disorder realizations.
           The inset in (b) shows the relative deviation
           for the marked point in the crossover regime (dashed line) from two heuristic interpolating laws
           between WD and Poisson: Semi-Poisson (grey) \cite{SP} and Berry-Robnik \cite{PR1993} (black).  
           As expected \cite{SP,PR1993}, correspondence is not perfect with neither of those laws, but the 
           Berry-Robnik lies closer overall. 
    }
  \label{fig:2}
\end{figure}

Our results of Figs.~\ref{fig:1} and \ref{fig:2} depend on the
procedure chosen to unfold the energy spectrum (i.e., to make the density of states approximately constant), 
which is necessary to compare to the normalized theoretical predictions \cite{KB2004,MPBS1993}. 
We used a rescaling of the levels by the numerically obtained local density of the raw data, 
which is independent of further assumptions on the original level density.
Choosing a small window (over $5\ldots 10$ levels) for computing the local
average permits an optimal match on smaller scales, whilst larger windows (over $20\ldots 40$ levels) 
are chosen to compute the number variance. 
Our analysis considered all levels of a given spectrum at fixed parameters, and 
we checked that excluding levels at the band edges does not change our results, as long as we
stay in the dilute limit of $UN/L \lesssim 1$. \\
For $\epsilon \lesssim 0.5$ and $\epsilon \gg 1$, we observed a trend towards globally regular 
dynamics, a consequence of good, yet not perfect Poisson statistics. In the limit $\epsilon \to 0$, the various QM blocks
uncouple, and as exercised in \cite{BK2003,KB2004,TMW2007} one has to concentrate just on
one of these blocks in the spectral analysis. Any small $\epsilon \neq 0$, however, destroys the translation
invariance, making an analysis of the {\it full} spectrum quite intricate. 
We, therefore, concentrate for a moment on the case of large disorder. As shown in Fig.~\ref{fig:1}(c), 
this limit is well characterized by a Poisson distribution. This result is expected, 
since the eigenstates become pinned at the randomly 
distributed minima of the potential, leading to a small residual overlap between them. 
Of course, for finite $U$ and $L$, such a localization
cannot be perfect \cite{SBZ1991}. Indeed, we observe a better correspondence in our
system with the Poisson prediction (see inset of Fig. \ref{fig:1}(c)) for smaller fillings, 
consistent with single-particle localization theory \cite{FFGP1985}. This trend was observed
when increasing $L < 20$ at fixed $N=3$ and $4$, or when decreasing $1 < N \leq 4$ 
at fixed $10 < L < 20$. \\
The DLT occurs for all data sets shown in Fig.~\ref{fig:2}
at a critical value $\epsilon_{\rm cr} \approx 4$. We characterize this crossover by a
$\chi^2$ statistical test \cite{nr}, which measures the deviation from WD.
The saturation of our $\chi^2_{\rm WD}$ measure for large $\epsilon$, at the right of the graphs
in Fig.~\ref{fig:2}, signals the convergence to a Poisson distribution, an example of which is
explicitly shown in Fig.~\ref{fig:1}(c).
For fixed $N$ and $L$, $\epsilon_{\rm cr}$ scales linearly with $U=J$, as expected since our 
Hamiltonian in \eqref{eq:1} is scale invariant for a given realization of disorder (i.e., 
$U=J$ defines the energy scale for $\epsilon$). At fixed $U=J$,
the crossover to the localized regime depends on the filling, and our results from 
Fig.~\ref{fig:2} suggest the following functional form $\chi^2_{\rm WD}(x)$, 
with $x\equiv \epsilon \sqrt{L/N}$, in the range
$\epsilon \simeq 4 \ldots 15$. It is numerically hard to obtain a full
scaling function for interacting systems with larger $N$ and $L$, since it is necessary
to diagonalize the full system, not just one QM block as in \cite{BK2003,KB2004,TMW2007}.

Interestingly, our spectral analysis of Figs.~\ref{fig:1} and \ref{fig:2} does not show a dependence on $\theta$
in the localized regime ($\epsilon \gtrsim 4$), whilst especially for
$\epsilon \lesssim 1$ (see inset of Fig.~\ref{fig:2}(a)) both couplings by $\theta \neq 0$ and 
by $\epsilon \neq 0$ can conspire to enhance the quantum chaotic properties of the {\it full} spectrum 
(i.e., not only of a subblock of fixed QM). We therefore can use, to some extent,
both parameters as independent handles to change the {\it global} spectral properties.

\begin{figure}
  \includegraphics [width=\linewidth]{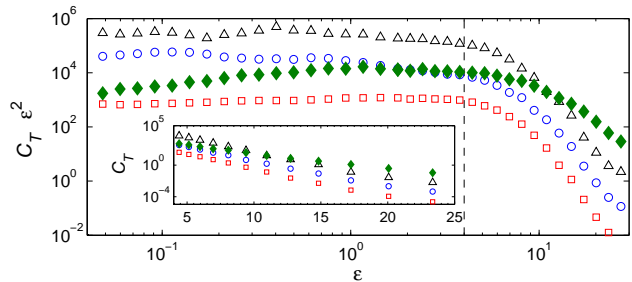}
  \caption{(color online) Level curvature $C_T$ as a function of the disorder strength for 
           $N=2$ ($\Box$), $3$ ($\circ$), $4$ ($\triangle$) and $J=1=U, \theta=0, L=15$. 
           Two different scalings are chosen to highlight the DLT around   
           the dashed line. Increasing the filling
           (particularly for $N=4,L=10$, shown as full diamonds) leads to deviations from the 
           single-particle scalings for both small and large disorder. 
    }
  \label{fig:3}
\end{figure}

\begin{figure}
  \centerline{\epsfig{figure=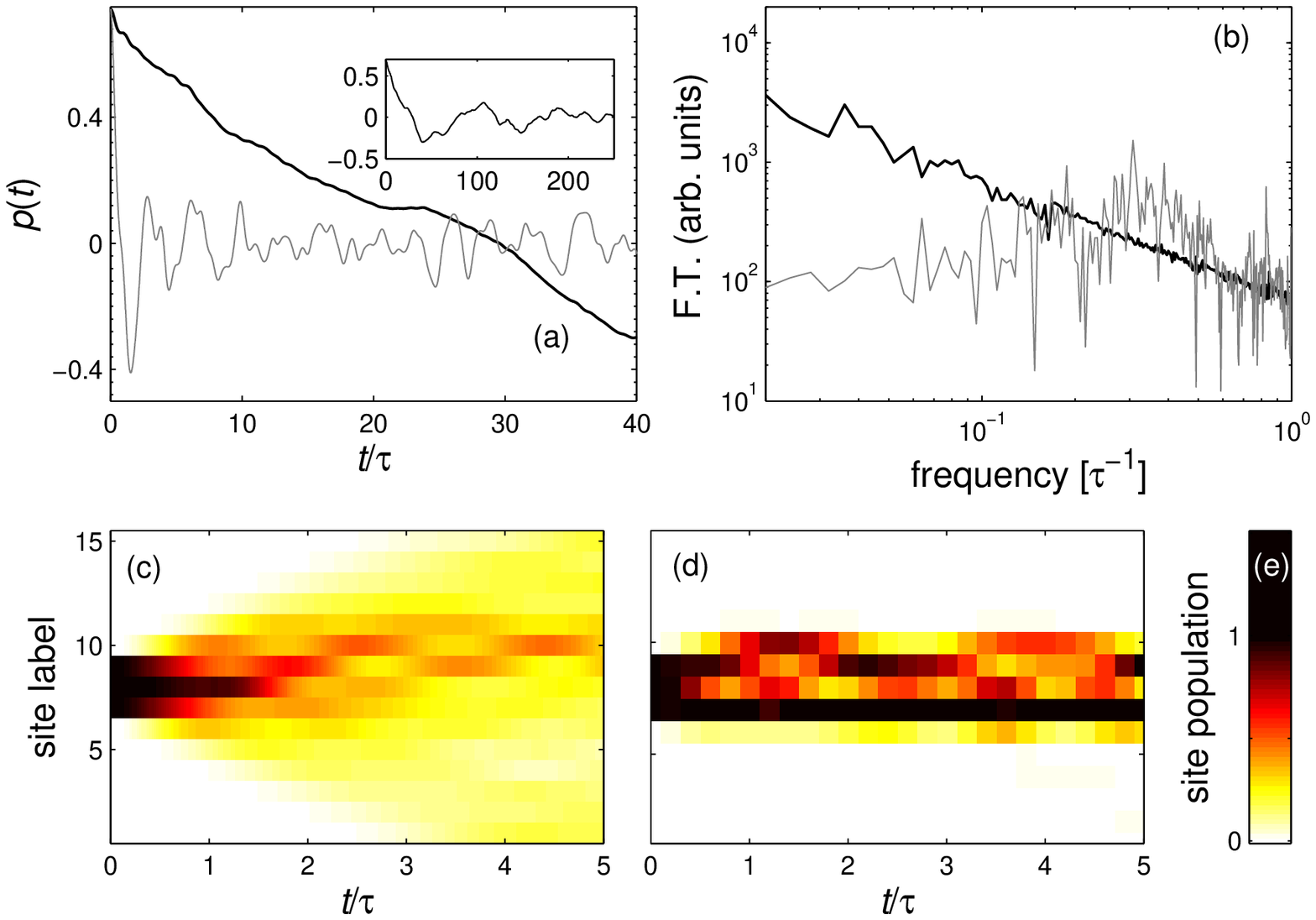,width=\linewidth}} 
  \caption{    (color online)
               (a) $p(t)$ for a three particle initial state with
               fixed QM, for $L=15, J=1=U, \theta=0$, and (b) Fourier transforms
               of the curves from (a), for $\epsilon=1$ (thick lines) and $\epsilon=10$ (thin grey lines).
	       The evolution of the well populations for a box initial state
               of $N=3$ atoms in real space is shown in (c) for $\epsilon=2$ and in (d) for $\epsilon=10$
               (same parameters as in (a), color code defined in (e)). 
               The time unit is $\tau = 2\pi\hbar/J[E_R] \approx 15 \rm \ ms$ \cite{BK2003}, 
               for a lattice constant of the order $\rm 412 \rm \ nm$ and a lattice depth of 10 recoil energies $E_R$ 
               for $^{87}Rb$ atoms \cite{Bloch2005,BEC_mott}. The inset in (a) highlights the slow decay for $\epsilon=1$.
    }
  \label{fig:4}
\end{figure}

The dependence of the spectrum on the choice of the boundary conditions defined
by $\theta$ is reminiscent of the Thouless conductance, another prominent measure
to characterize the transition between extended and localized states \cite{KM1993}.
We computed the Thouless conductance, which essentially is given by the curvature
$C_T \equiv \langle |d^2E/d\theta^2|\rangle 
\approx \langle |2[E(\theta)-E(0)]|/\theta^2 \rangle$, for $\theta \to 0$ \cite{CGIMZ1994,RB2003}, which
was geometrically averaged \cite{CGIMZ1994} over the full spectrum and 40 realizations of disorder.
Our results are shown in Fig.~\ref{fig:3}. In contrast to the distributions of nearest level spacings,
where the small but finite mixing of QM blocks at $\epsilon \to 0$ does not allow to well 
characterize the true type of dynamical regime, the curvature is a {\it local} property of the spectrum. 
Hence, in the diffusive limit $\epsilon \to 0$, we find the expected divergence
$C_T \propto \epsilon^{-\alpha}$, with $\alpha \approx 2 \ldots 1.8$ ($N=2\ldots 4, L=15$)
and $\alpha \approx 1.3$ ($N=4, L=10$).
In analogy to Fig.~\ref{fig:2}, the crossover between the diffusive (quantum chaotic) and
the localized regime sets in at $\epsilon_{cr} \approx 4$. For $\epsilon \gtrsim 4$, our
results confirm an exponential scaling (typical of finite-size localized systems) 
$C_T \propto \exp( - {\rm const}\cdot \epsilon^\beta )$, with $\beta \approx 0.8$ ($N=2\ldots 4, L=15$)
and $\beta \approx 0.5$ ($N=4, L=10$). The systematic deviation of both exponents from the 
single-particle predictions ($\alpha = 2$ and $\beta = 1$ \cite{CGIMZ1994}) with increasing
filling factor highlights the structural change of the level dynamics in the presence of 
atom-atom interactions.\\
In the following, we focus on experimentally observable consequences of the spectral analysis presented
so far. Fig.~\ref{fig:4} shows the temporal evolution of initial states, typically prepared in experiments
with Bose-Einstein condensates. Panel (a) presents the mean momentum on the lattice 
\cite{BK2003,RB2003,K2006}, defined as
\begin{eqnarray}
\label{eq:4}
p(t) &\equiv &\frac{1}{2\ii N} \bra{\psi(t)} \sum_{\ell} \left( a_\ell^\dag a_{\ell-1} - h.c. \right) \ket{\psi(t)} 
\nonumber \\
&=& \frac{1}{N} \bra{\psi(t)} \sum_{\kappa} \sin(\frac{2\pi \kappa }{L}) b_\kappa^\dag b_{\kappa} \ket{\psi(t)}\;,
\end{eqnarray}
in the direct and reciprocal space, respectively, for 
$\ket{\psi~(~t~=~0~)} = (b_{\kappa}^\dag)^N \ket{0}/\sqrt{N!}$, with
$\kappa = 2 $. For strong disorder, the momentum decays almost instantaneously to zero, and is further
characterized by small, random fluctuations. The quantum chaotic behavior for $\epsilon=1$ is visible in the
strongly correlated large-scale fluctuations, characterized by a slowly, algebraically
decaying Fourier transform of the time series $p(t)$, c.f.  Fig.~\ref{fig:4}(b). The latter implies a large number of
modes being present in the evolution of $p(t)$, a standard signature of complex dynamics \cite{BS1988}.
Complex transport behavior was predicted also in \cite{wang2004} by analyzing the power spectrum of oscillations
in a three-well system. In contrast to (a,b), Figs.~\ref{fig:4}(c,d) show the real space dynamics of a box-distributed 
initial state with one atom in wells $\ell = 7,8,9$ and none elsewhere. Whilst the chaotic mixing of all wells dominates for
$\epsilon=2$, in the localized regime $\epsilon=10$ we observe the expected strong pinning at the random minima
of the potential (particularly in the 7th well in Fig.~\ref{fig:4}(d)). The fluctuations in the latter case arise from 
our finite values of $U$ and $L$. For a typical realization of $\{\epsilon_\ell\}$, the evolution will be asymmetric
as seen in (c), which allows one to distinguish it from the $\epsilon=0$ case. Equivalently, whilst for
$\epsilon=0$ QM is conserved, for $\epsilon \neq 0$ QM starts to deviate from its initial value, which is
$0$ for the data in (c,d). Fig.~\ref{fig:4} presents the limits of fully developed global quantum chaos 
(without any residual symmetry in the system) and strong localization. Yet, the crossover between the two
regimes is systematic, and as exemplified, even the evolution of single, reduced experimental observables of 
our many-body problem can be used to directly visualize the change in the quantum spectra analyzed 
in Figs.~\ref{fig:1}-\ref{fig:3}. Of experimental relevance is in particular the difference in the 
short-time evolution of $p(t)$ for experimental detection times $< 1\rm \,s$. \\
In summary, we showed how one can scan between the different dynamical regimes of the Bose-Hubbard system, 
characterized by global quantum chaos and by essentially localized bosons, by
varying the strength of static disorder. The dynamics of initial states which are far from eigenstates
of the system could be used as a clear experimental signature of this crossover. Moreover, our analysis
of the many-body level curvatures opens a new link to transport problems in mesoscopic solids 
\cite{CGIMZ1994,KM1993} and photonic lattices \cite{SBFS2007}, where global chaotic
properties are accessible by conductance measurements.
As adumbrated in \cite{WPI1997}, the here presented, extended and unifying characterization of the spectral 
properties of a disordered many-body problem, may be useful to obtain, for instance, experimentally accessible 
estimates for the localization properties for such complex systems. 
However, the necessary scaling arguments \cite{KM1993} as a function of 
$N$ and $L$ make such an approach challenging for up-to-date computational resources. 

PB acknowledges support by the {\it Lagrange Project}-CRT Foundation and SW within the
Excellence Initiative by the DFG through the Heidelberg Graduate School of Fundamental Physics
(grant N$^{o}$ GSC 129/1). We are grateful to A. Montorsi, V. Penna, P. Schlagheck, and G. Veble for useful discussions.


\begin{thebibliography}{99}

\bibitem{Bloch2005}
O. Morsch and M. Oberthaler, \rmp {\bf 78}, 179 (2006);
I. Bloch {\it et al.}, preprint arXiv:0704.3011.

\bibitem{Paul2007}
see, e.g., T. Paul {\it et al.}, \prl {\bf 98}, 210602 (2007);
B. Shapiro, \prl {\bf 99}, 060602 (2007).

\bibitem{Lye2005}
J. E. Lye {\it et al.}, \prl {\bf 95}, 070401 (2005);
D. Cl\'ement {\it et al.}, {\it ibid.} 170409;
C. Ford {\it et al.}, {\it ibid.} 170410;
T. Schulte {\it et al.}, {\it ibid.} 170411;
L. Fallani {\it et al.}, {\it ibid.} {\bf 98}, 130404 (2007).

\bibitem{BEC_mott}
M. Greiner {\it et al.}, Nature (London) {\bf 415}, 39 (2002);
T. St\"oferle {\it et al.}, \prl {\bf 92}, 130403 (2004).

\bibitem{Ess2005}
M. K\"ohl {\em et al.}, Phys. Rev. Lett. {\bf 94}, 080403 (2005).

\bibitem{TMW2007}
A. Tomadin {\it et al.}, \prl {\bf 98},  130402 (2007).

\bibitem{BK2003}
A. Buchleitner and A. R. Kolovsky, \prl {\bf 91}, 253002 (2003).

\bibitem{KB2004}
A. R. Kolovsky and A. Buchleitner, \epl {\bf 68}, 632 (2004).

\bibitem{WPI1997}
D. Weinmann {\it et al.}, \jpf {\bf 7}, 1559 (1997).

\bibitem{MPBS1993}
G. Montambaux {\it et al.}, \prl {\bf 70}, 497 (1993).

\bibitem{CGIMZ1994}
G. Casati {\it et al.}, \prl {\bf 72}, 2697 (1994);
D. Braun {\it et al.}, \prb {\bf 55}, 7557 (1997).

\bibitem{KM1993}
B. Kramer and A. MacKinnon, Rep. Prog. Phys. {\bf 56},  1469  (1993).

\bibitem{RB2003}
B. S. Shastry and B. Sutherland, \prl {\bf 65}, 243 (1990);
R. Roth and K. Burnett, \pra {\bf 67}, 031602(R) (2003).

\bibitem{SBZ1991}
R. T. Scalettar {\it et al.}, \prl {\bf 66}, 3144 (1991).


\bibitem{phase}
see, e.g., A.~M. Rey {\it et al.} \pra {\bf 75}, 063616 (2007).

\bibitem{AOC2005}
L. Amico {\it et al.}, \prl {\bf 95}, 063201 (2005).

\bibitem{MG2004}
B. Mieck and R. Graham, \jpa {\bf 37}, L581 (2004);
E. Courtade {\it et al.}, \pra {\bf 74}, 031403(R) (2006).

\bibitem{BPV2005}
P. Buonsante {\it et al.}, \pra {\bf 72}, 043620 (2005).

\bibitem{K2006}
A. R. Kolovsky, New J. Phys. {\bf 8}, 197 (2006).

\bibitem{FFGP1985}
M. Feingold {\it et al.}, \prb {\bf 31}, 6852 (1985). 

\bibitem{SP}
S. N. Evangelou and J.-L. Pichard, \prl {\bf 84}, 1643 (2000).

\bibitem{PR1993}
T. Prosen and M. Robnik, \jpa {\bf 26}, 2371 (1993).

\bibitem{nr}
W. H. Press {\it et al.}, {\it Numerical Recipes} (Cambridge University Press, Cambridge 1993).

\bibitem{BS1988}
See, e.g., L. E. Reichl, {\it The Transition to Chaos} (Springer, 2004);
R. Bl\"umel and U. Smilansky, \prl {\bf 60}, 477 (1988).

\bibitem{wang2004}
D.-W. Wang {\it et al.}, \prl {\bf 92}, 076802 (2004).

\bibitem{SBFS2007} 
T. Schwartz {\it et al.}, Nature (London) {\bf 446}, 52 (2007).

\end{thebibliography}
\end{document}